\documentclass[11pt,twocolumn]{article}

\usepackage[utf8]{inputenc}
\usepackage[a4paper, top=3.5cm, bottom=3.5cm]{geometry}
\usepackage[T1]{fontenc}
\usepackage{newtxtext,newtxmath}  
\usepackage{helvet}
\usepackage{courier}
\usepackage[hyphens]{url}
\usepackage{graphicx}

\usepackage{amsmath, amssymb, amsfonts}
\usepackage{xcolor}
\usepackage{booktabs}
\usepackage{caption}
\usepackage{paralist}
\usepackage{algorithm}
\usepackage{algorithmic}
\usepackage[most]{tcolorbox}   
\usepackage{hyperref}

\definecolor{jacopocolor}{HTML}{BF092F}
\definecolor{matteocolor}{HTML}{E5BA41}
\definecolor{elcolor}{HTML}{3B9797}    
\definecolor{enemmicolor}{HTML}{1261A0} 

\providecommand{\ie}{\emph{i.e.,} }

\newcommand{\citebanner}{%
  \begin{tcolorbox}[colback=yellow!12, colframe=black!70, boxrule=0.5pt,
                    arc=2mm, left=5pt, right=5pt, top=3pt, bottom=3pt,
                    halign=justify, fontupper=\small\normalfont]
  This paper is an earlier version of a published work. If you wish to cite it,
  please refer to the following instead:
  \begin{itemize}\setlength{\itemsep}{0pt}\setlength{\topsep}{2pt}
    \item Nudo, Jacopo, et al. ``On the Vulnerability of Community Notes to
    Participation Concentration.'' In: \emph{Computer Security -- ESORICS 2026},
    Lecture Notes in Computer Science, vol.~16969, pp.~482--501.
    Springer, Cham (2027).
    \href{https://doi.org/10.1007/978-3-032-38695-3_24}{doi:10.1007/978-3-032-38695-3\_24}
  \end{itemize}
  \end{tcolorbox}}

\title{\citebanner\vspace{1.2em}Hyperactive Minority Alters the Stability of Community Notes}

\author{
    Jacopo Nudo\thanks{Sapienza University of Rome, Department of Computer Science, Rome, Italy. Email: jacopo.nudo@uniroma1.it} \and
    Eugenio Nerio Nemmi\footnotemark[1] \and
    Edoardo Loru\thanks{Department of Computer, Control and Management Engineering, Sapienza University of Rome, Rome, Italy} \and
    Alessandro Mei\footnotemark[1] \and
    Walter Quattrociocchi\footnotemark[1] \and
    Matteo Cinelli\footnotemark[1]
}

\date{}

\begin{document}

\maketitle

\begin{abstract}
As platforms increasingly scale down professional fact-checking, community-based alternatives are promoted as more transparent and democratic. The main substitute being proposed is community-based contextualization, most notably Community Notes on X, where users write annotations and collectively rate their helpfulness under a consensus-oriented algorithm. This shift raises a basic empirical question: to what extent do users’ social dynamics affect the emergence of Community Notes?
We address this question by characterizing participation and political behavior, using the full public release of notes and ratings (between 2021 and 2025). We show that contribution activity is highly concentrated: a small minority of users accounts for a disproportionate share of ratings. Crucially, these high-activity contributors are not neutral volunteers: they are selective in the content they engage with and substantially more politically polarized than the overall contributor population. 
We replicate the notes' emergence process by integrating the open-source implementation of the Community Notes consensus algorithm used in production. This enables us to conduct counterfactual simulations that modify the display status of notes by varying the pool of raters. Our results reveal that the system is structurally unstable: the emergence and visibility of notes often depend on the behavior of a few dozen highly active users, and even minor perturbations in their participation can lead to markedly different outcomes. In sum, rather than decentralizing epistemic authority, community-based fact-checking on X reconfigures it, concentrating substantial power in the hands of a small, polarized group of highly active contributors.
\end{abstract}

\section{Introduction}

Social media platforms have altered how information is spread and consumed \cite{bakshy2012role, del2016spreading,sharot2020people}, shifting away from the top-down model of traditional media toward decentralized, bottom-up dynamics. In this environment, algorithmic curation leads individuals to consume content that aligns with their pre-existing beliefs \cite{bakshy2015exposure, del2017modeling}, fostering the creation of ideologically homogeneous groups \cite{cinelli2021echo} where misinformation can thrive. As a consequence, limiting the diffusion of false or misleading content has become a central challenge for both platforms and researchers \cite{del2016spreading, lazer2018science,theSpread}
Early responses relied heavily on professional fact-checking organizations, tasked with identifying and labeling inaccurate content. However, this approach has faced two structural limitations. First, it has raised political and normative debates about who holds the authority to define truth in increasingly polarized information environments \cite{ persily2020social, allen2021scaling}. 
Second, expert-based fact-checking has proven difficult to scale, due to its reliance on trained professionals and the volume and velocity of online content \cite{zollo2017debunking,allen2021scaling}.

While automated tools for  misinformation detection have shown promising results \cite{shu2017fake, perez2018automatic, dixon2018measuring}, they cannot fully replace human epistemic judgment. Against this backdrop, some platforms have increasingly turned to community-based approaches, framing them as more transparent, scalable, and democratically legitimate alternatives to centralized fact-checking \cite{apNewsCrowd-sourced}.
In early 2021, Twitter introduced Birdwatch, a community-driven initiative that allowed users to add contextual notes to tweets suspected of being misleading \cite{wojcik2022birdwatch}. Participants can write explanatory annotations, which are then rated by other contributors for helpfulness, producing a decentralized and consensus-based system for contextualizing content. In 2022, Birdwatch was rebranded as Community Notes (CNs) following Twitter’s transition to X and continuing under the same core principles \cite{xIntroduction}. CNs operates as a parallel, selective layer to the main platform, where a subset of users collaboratively monitors and annotates content.
This model has since been adopted by other platforms. Meta launched a similar community-based system as part of its strategy to move beyond traditional fact-checking while emphasizing free expression \cite{metaIntroducingCommunity, fbMoreSpeech}, and TikTok introduced its Footnotes program, reporting tens of thousands of contributors within its first year \cite{tiktokTestingFeature}.
Despite the growing interest in the effectiveness of community-driven systems, such as CNs, in reducing the spread of misinformation \cite{chuai2023roll,slaughter2025community}, the social dynamics that shape these systems have received comparatively less attention. In particular, by allowing contributors to actively choose which content to evaluate and to rate the contributions of others, community-based fact-checking may replicate and be affected by certain dynamics observed on social media \cite{nogara2022disinformation,avalle2024persistent,nudo_size}, namely the extreme concentration of activity and polarization dynamics \cite{del2016echo,cota2019quantifying,cinelli2020covid,gonzalez2023asymmetric}.

\subsection{Contributions}
In this paper, we provide a large-scale empirical characterization of CNs as a socio-technical system for community-based fact-checking. Leveraging the complete public record of notes and ratings from 2021 to 2025, together with the open-source implementation of the consensus algorithm, we analyze how participation patterns and political behavior shape the system’s outcomes using counterfactual simulations.

First, we show that contribution and rating activity in CNs is highly unequal. Consistent with activity patterns observed in other social media platforms \cite{nogara2022disinformation, avalle2024persistent}, a small minority of users generate a disproportionately large share of ratings, concentrating influence over note outcomes in the hands of a few. This concentration grows over time and stabilizes after an initial adoption period, suggesting it is a structural feature rather than a transient phenomenon. Moreover, raters display selectivity, i.e., they repeatedly focus on tweets authored by profiles for which they have already expressed a rating.

Second, we demonstrate that activity concentration is not politically neutral. The most active contributors are significantly more polarized than the overall population of participants, exhibiting stronger and more asymmetric political leanings. Given that highly active users are both selective and polarized, this leads to a rating process that can be influenced by political asymmetry.

Third, we connect these participation asymmetries to the expected output of the CNs consensus algorithm. Although the algorithm is explicitly designed to surface notes only when broad agreement across political groups is achieved, we show that high levels of activity concentration among top contributors, who are also the most polarized, can meaningfully influence which notes emerge. Through counterfactual simulations enabled by access to the algorithm code, we assess how changes in the composition of contributors affect note visibility. These simulations reveal that excluding even a small fraction of highly active users can substantially alter the set of notes that reach public visibility.

Taken together, our findings reveal that community-driven moderation mirrors key characteristics of online social interactions, including activity concentration, uneven participation, and political polarization. These results ultimately underscore how large-scale online social dynamics can undermine the effectiveness of crowd-sourced moderation systems as a fact-checking mechanism on social media.

\subsection{Data \& Methodology}
\label{subsec:methods}

\begin{figure}[h]
  \centering
  \includegraphics[width=0.47\textwidth]{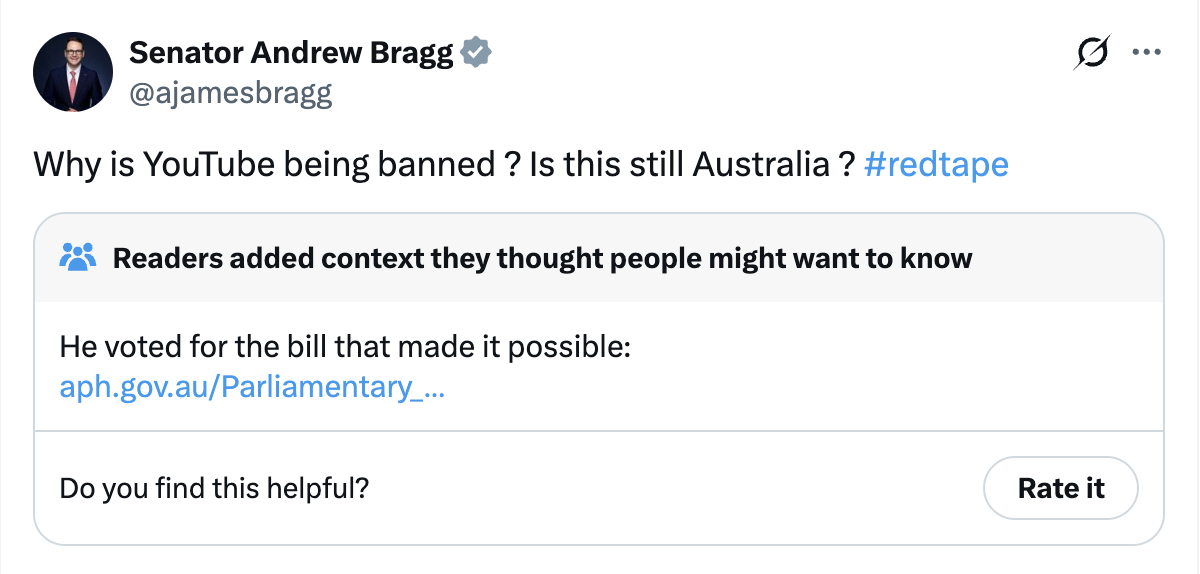}
  \caption{Example of how a Community Notes labeled as \texttt{Helpful} is presented to all the users on X, including the contextual explanation and the associated sources displayed beneath the original tweet.}
  \label{fig:crhnotes_fun}
\end{figure}

Community Notes is a collaborative system that allows users, referred to from now on as \textit{contributors}, to propose textual annotations, referred to as \textit{notes}, in response to individual tweets. 
The purpose of a note is to add contextual information to a tweet, including additional sources supporting the note's claim. 
An X user, in order to become a member of Community Notes, must meet a set of eligibility requirements, including maintaining an account that has been active for at least six months, having a verified phone number, and no recent violations of the platform’s policies; once admitted, the user initially participates as a \textit{raters}.
If, over time, their behavior is deemed reliable, the platform allows them to propose notes under tweets, thereby becoming  \textit{contributors}.
Once a note is proposed, all \textit{raters} can see it under the associated tweet, and can provide a rating chosen from: \texttt{Helpful}, \texttt{Somewhat Helpful}, or \texttt{Not Helpful}.
As a note collects rating events, the Community Notes algorithm periodically evaluates the level of agreement reached among raters with historically diverse rating behavior (\ie raters who have disagreed in the past). This broad agreement is referred to as consensus. 
Upon publication, a note is initially assigned the status \texttt{Needs More Ratings} (NMR), and it is only visible among members of community notes (all the \texttt{raters}). As ratings are submitted, if the they meet the algorithmic criteria for sufficient cross-partisan consensus, note's status is updated accordingly. Notes that are deemed helpful are labeled \texttt{Currently Rated Helpful} (CRH), whereas notes for which the consensus indicates a lack of helpfulness are labeled \texttt{Currently Rated Not Helpful} (CRNH). If no consensus is reached, the note retains its original NMR status.
These status assignments determine whether a note remains visible only to contributors or becomes publicly visible on the platform.
Notes labeled CRH are publicly displayed beneath the associated tweet (see Fig.~\ref{fig:crhnotes_fun}), whereas notes labeled CRNH substantially reduces their visibility within the platform, requiring explicit user action to view them.

We rely on the complete official Community Notes dataset released by X\footnote{\url{https://communitynotes.x.com/guide/en/under-the-hood/download-data}}, spanning January 28, 2021, to December 2, 2025. The dataset contains over 2.2 million submitted notes, approximately 1.3 million distinct contributors who have rated at least one note, and a total of approximately 184 million rating events. While the official CNs dataset provides a unique identifier for each note and for the associated tweet, it does not include the identifier of the tweet's author. Therefore, for the analyses explicitly concerned with polarization and selective engagement, which would require the authors' identifier, we focus on a politically annotated sample of notes constructed by \cite{renault2025republicans}\footnote{\url{https://github.com/trenault/CommunityNotes}}. 
This subset includes only notes referring to tweets authored by accounts for which a partisan label is inferred, yielding 218{,}382 notes associated with 162{,}228 tweets from 18{,}251 Republican and 20{,}889 Democratic profiles. By joining the ratings dataset with this sample via the Community Notes identifier, we can link contributors and their ratings to the corresponding tweet author. The resulting dataset comprises 18{,}248{,}161 rating events (about 10\% of the total), contributed by 733{,}750 distinct raters, providing a politically interpretable space in which to study partisan asymmetries and selectivity.
To reproduce platform behavior in terms of notes' ranking as faithfully as possible, we integrate the open-source official implementation of the CNs consensus algorithm used in production\footnote{\url{https://github.com/twitter/communitynotes}}. To run the algorithm and test alternative counterfactual scenarios, we directly manipulate its two core input datasets: (i) the \textit{ratings} matrix, which encodes for each rater–note pair the rating expressed by the rater toward a given note (\texttt{Helpful}, \texttt{SomewhatHelpful}, or \texttt{NotHelpful}), and (ii) the \textit{noteStatusHistory}, which records the sequence of statuses that each note assumes over time (\texttt{NeedsMoreRatings}, \texttt{CurrentlyRatedHelpful}, or \texttt{CurrentlyRatedNotHelpful}). This allows us to replicate the decision logic governing note display and to run counterfactual simulations altering the composition of raters to evaluate how such changes influence the emergence of a note under a tweet.

\section{Related Work}
\label{related}
CNs represent a rare case in which a large-scale fact-checking system is both fully deployed in a production platform and openly accessible for scientific investigation. Beyond enabling users to collaboratively annotate and add context to potentially misleading content, it provides researchers with access to user-generated data, the full history of notes and ratings, and the underlying consensus algorithm. This combination makes CNs a particularly valuable setting for observing how disagreement and consensus emerge in a real-world community-driven moderation process, and how these dynamics shape the visibility of corrective information.
While several community-based fact-checking systems exist, CNs is the most extensively studied in the literature due to its transparency and the accessibility of its data. Consequently, in the following sections we focus exclusively on CNs when discussing collaborative annotation, disagreement, and consensus dynamics.
A growing body of literature has examined whether this form of collective evaluation can effectively limit the diffusion of misinformation. Several studies report that crowd-sourced approaches to fact-checking are scalable and cost-effective, and can achieve levels of accuracy comparable to expert-based models \cite{slaughter2025community,martel2024crowds,saeed2022crowdsourced,la2024crowdsourced}. These results have positioned CNs as a promising alternative, or complement, to professional fact-checking mechanisms, particularly in contexts where centralized fact-checking is contested or difficult to scale.
Beyond accuracy, perceived credibility is a key factor in the effectiveness of corrective interventions. Research shows that notes created by the community are generally considered more trustworthy than labels from a centralized third-party fact-checker, across the political spectrum \cite{drolsbach2024community}.
Several studies highlight that CNs do not operate independently from traditional fact-checking ecosystems. Contributors frequently rely on professional fact-checking sources when writing notes \cite{borenstein2025can,kangur2024checks}, suggesting that community-based systems often function as hybrid mechanisms rather than fully independent alternatives. Even when platforms do not explicitly integrate professional fact-checkers, community-based fact-checking systems may nonetheless benefit indirectly from their presence within the broader information environment \cite{pilarski2024community}.
Related work also suggests that exposure to community-authored corrections is less likely to trigger disengagement or reactance compared to individual misinformation flags, indicating a potentially lower polarizing effect \cite{kim2025differential}. Together, these findings support community-based moderation as both legitimate and socially acceptable.
A recurring theme in the literature concerns the role of timing. While the presence of a CN generally reduces engagement with misleading content, its effectiveness is highly sensitive to publication delays \cite{truong2025delayed,brashier2021timing,pilarski2024community}. Missing the window of peak attention substantially weakens corrective effects \cite{slaughter2025community,chuai2023roll,borenstein2025can}, especially during sudden bursts of information diffusion such as infodemics \cite{cinelli2025infodemic}. In practice, this temporal constraint is severe: only a small fraction of submitted notes achieve “helpful” status and are published, with an average delay of over 24 hours \cite{mohammadi2025birdwatch}. As a result, most users never encounter CNs \cite{binder2023communitynotes}, and a large majority of misleading posts that have accurate notes are never shown corrective context \cite{counterhate}.
These limitations have prompted concerns about the scope and coverage of community-based moderation. Ambiguous content and under-recognized forms of harm may escape annotation altogether \cite{matamoros2025importance}, and highly influential users can remain insufficiently corrected, reducing the overall effectiveness of the system \cite{prollochs2022community}. More recent work further shows that community-based fact-checking is not immune to political asymmetries. Analyses of CNs reveal that Republican posts are significantly more likely than Democratic posts to be flagged as misleading, a pattern that cannot be explained by rater bias or user base composition alone \cite{renault2025republicans}. 
In a polarized environment, contributors may be driven more by the desire to correct ideological opponents than by neutral accuracy assessment, potentially reflecting or amplifying partisan dynamics \cite{augenstein2025community,yasseri2023can,allen2022birds}. In addition, participation itself is fragile: contributors whose notes are rarely published are less likely to remain active, raising concerns about the long-term sustainability of engagement in such systems \cite{arjmandi2025threats}.
A study uses synthetic data to systematically evaluate how the Community Notes algorithm performs \cite{truong2025community}, measuring how often helpful and unhelpful notes are correctly published. It finds that the system is highly vulnerable to rater biases and strategic manipulation, where even a small minority of bad actors can suppress reliable information, raising concerns about the reliability of the process.
Finally, recent research has explored the use of large language models (LLMs) to support CNs, for example, by assisting with note drafting or early detection of emerging events \cite{li2025scaling}. While these approaches aim to reduce response times and improve coverage, evidence suggests that algorithmic assistance alone is insufficient. Producing notes that foster agreement across politically diverse users appears to require the combination of LLM-based candidate generation with consensus-driven scoring mechanisms, rather than replacing human judgment \cite{de2025supernotes}.

\section{Analysis}
We analyze Community Notes along three dimensions:
\begin{inparaenum}[(i)]
    \item the concentration of rating activity,
    \item the political polarization of highly active raters,
    \item the sensitivity of the algorithm governing which notes are displayed to the participation of hyperactive raters.
\end{inparaenum}

\subsection{Concentration of activity}

\begin{figure}[h]
  \centering
  \includegraphics[width=0.5\textwidth]{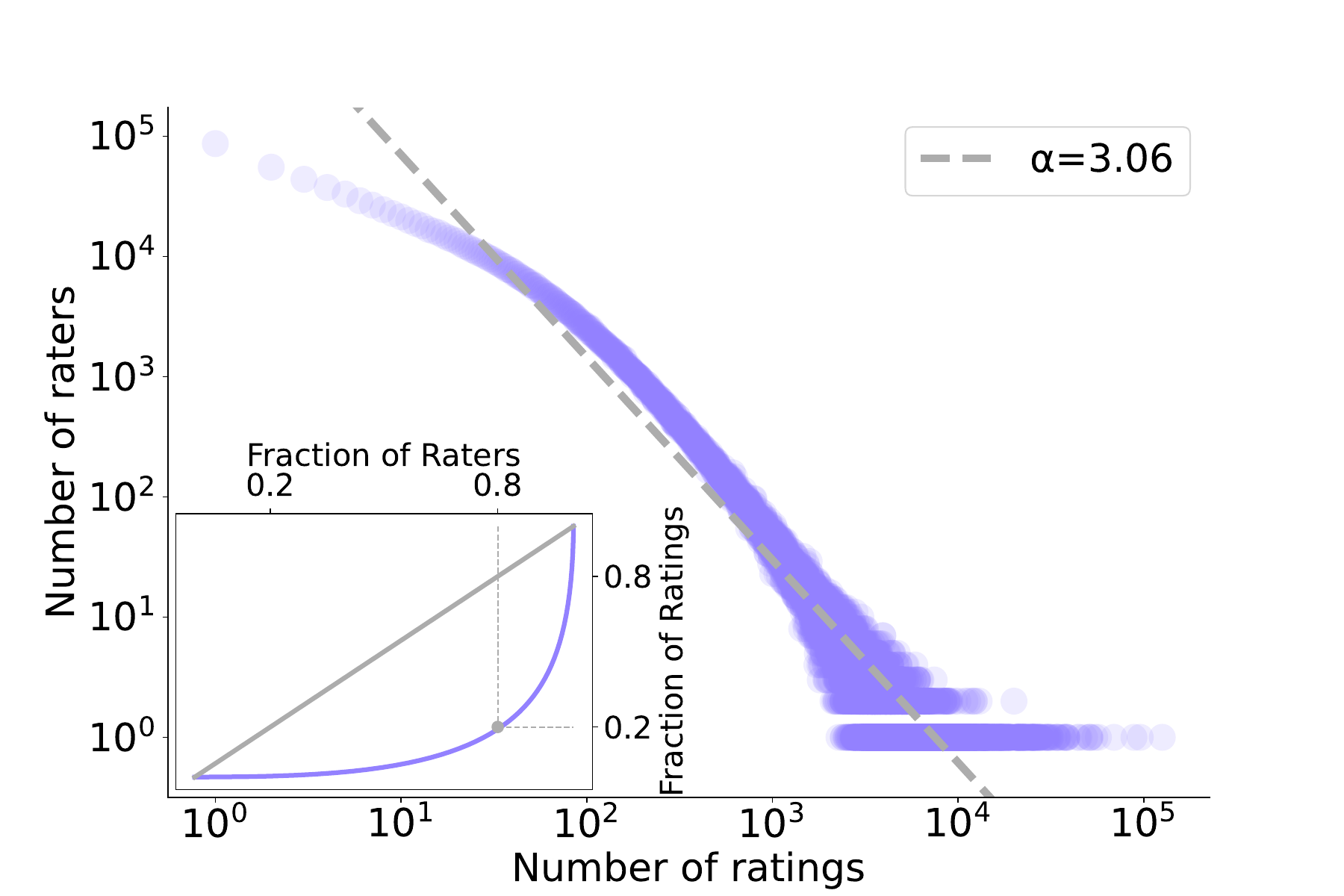}
  \caption{Log-log distribution of ratings per contributor, with an inset showing the Lorenz curve. The heavy-tailed shape indicates that most users give few ratings, while a small minority is responsible for the majority of contributions: 20\% of users account for roughly 80\% of all ratings as displayed by the marker on the Lorenz curve.}
  \label{fig:ratings}
\end{figure}

In this section, we investigate how users who rate other contributors’ notes distribute their activity. First, we examine whether most users contribute uniformly, or if there exists a minority of highly active users responsible for most ratings. Next, we assess whether raters provide ratings broadly across all available notes or concentrate their activity on a smaller subset. 
In a consensus-driven fact-checking system, designed to be democratic and transparent, it would be desirable for all contributors to engage evenly and widely, as this can promote diverse perspectives and more heterogeneous judgment. By contrast, when evaluative activity is concentrated among a small group of users, the expected diversity of judgments decreases, and rating behavior increasingly mirrors heterogeneous dynamics such as engagement on social media. 
In Figure~\ref{fig:ratings}, we display how many users have contributed a given number of ratings. The resulting distribution appears heavy-tailed, indicating that most users only give a few ratings, while a smaller subset is highly active.
Following the methodology proposed in a previous study, the distribution of the number of ratings per user was modeled using a discrete power-law distribution \cite{clauset2009power}.
Specifically, for $x \ge x_{\min}$, the probability mass function can be expressed as
\begin{equation}
P(X = x) \propto x^{-\alpha},
\end{equation}
where $\alpha$ is the scaling exponent. 
Model parameters were estimated, yielding $\alpha = 3.06 \pm 0.04$ and a lower cutoff $x_{\min} = 4232$, determined by minimizing the Kolmogorov--Smirnov distance ($D = 0.013$).
A likelihood ratio test comparing the power-law fit with an exponential one strongly favors the power-law model ($R = 6.49$, $p < 10^{-4}$), providing clear evidence of statistically significant heavy-tailed behavior.
Furthermore, Figure~\ref{fig:ratings} includes an inset showing the corresponding Lorenz curve, computed as the cumulative share of ratings as a function of the cumulative share of raters.
In line with the Pareto principle (also known as the 80/20 rule) \cite{pareto1964cours}, the curve shows that approximately 20\% of contributors account for about 80\% of all ratings. 
To quantify this concentration, we compute the Gini coefficient \cite{gini1921measurement}:
\begin{equation}
G = \frac{\sum_{i=1}^{n}\sum_{j=1}^{n} |x_i - x_j|}{2n \sum_{i=1}^{n} x_i},
\end{equation}
where $x_i$ denotes the number of ratings produced by user $i$ and $n$ is the total number of users. The Gini coefficient is a measure in $[0,1]$ that quantifies inequality in a distribution, with higher values corresponding to higher inequality. 
Computing the coefficient over the entire observation window yields
\[
G = 0.78,
\]
indicating a high level of inequality in rating activity across users. The temporal evolution of the Gini coefficient is
\[
G_t = \{0.69,\, 0.78,\, 0.81,\, 0.79,\, 0.79\}, \quad t = 2021,\dots,2025.
\]
This sequence highlights a clear regime change: the coefficient increases markedly between 2021 and 2022, a period that coincides with the rapid expansion of CN adoption. In the subsequent years, the coefficient exhibits a stable pattern, suggesting that rating activity remains persistently concentrated within a small group of highly active contributors.

\begin{figure}[h]
  \centering
  \includegraphics[width=0.9\linewidth]{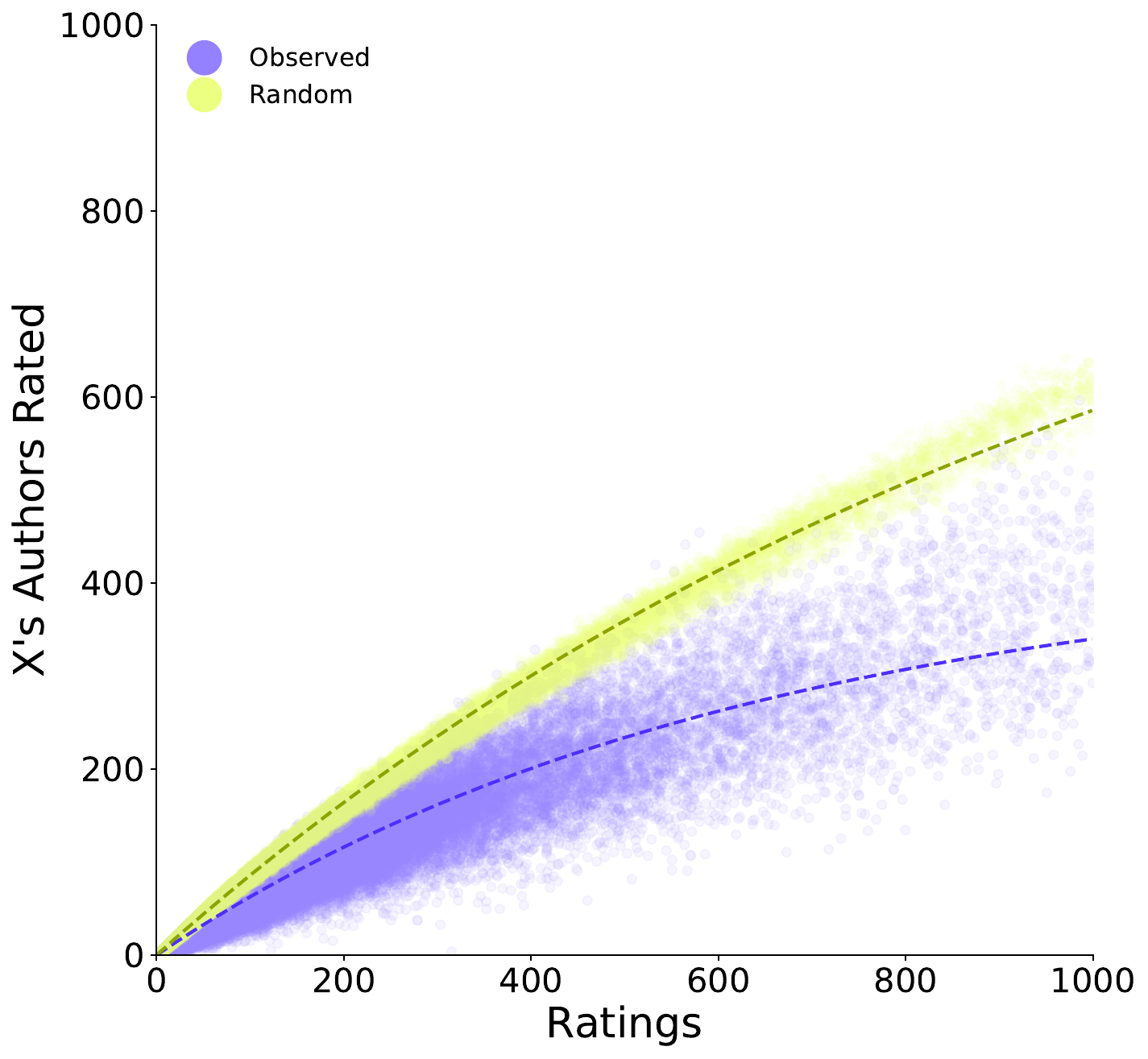}
  \caption{The plot shows the cumulative number of distinct authors rated as a function of the number of ratings provided by each rater. Observed data points are fitted with a saturation curve, indicating that raters concentrate their activity on a limited set of authors and reach the asymptotic number of engaged authors more quickly than expected under a random baseline obtained by reshuffling the author-rating couples.}
  \label{selective}
\end{figure}

Next, we investigate the extent to which individual raters distribute their activity uniformly across profiles subject to evaluation. 
For this analysis, which requires identifiers for both the raters and the authors of the tweets linked to the notes, we focus exclusively on the subset of ratings related to tweets authored by accounts associated with a political label, as defined in Section~\ref{subsec:methods}.
For each rater, we count both the total number of ratings provided and the number of distinct authors on X to whom these ratings referred. As illustrated in Figure~\ref{selective}, raters tend to concentrate their activity on a limited set of authors, repeatedly evaluating CNs written in response to tweets by the same set of authors. Furthermore, higher levels of rater activity are associated with greater concentration of ratings on specific tweet authors, indicating that selectivity intensifies as raters become more active. 
As the number of ratings exceeds the number of distinct tweet authors, some degree of concentration is expected even under random behavior. To account for this, we repeat the analysis after randomizing the authors of all rated tweets, obtaining a null baseline. As shown in Figure~\ref{selective}, the randomized curve exhibits limited saturation, but this effect is substantially weaker than in the empirical data, indicating a non-random preference in content selection. To better characterize this selectivity, we model both the empirical and randomized curves using the same saturation function, representing the cumulative number of distinct authors as a function of the number of ratings:
\begin{equation}
N(r) = N_{\text{asy}} \left( 1 - e^{-r / \tau} \right),
\end{equation}
where \(N(r)\) denotes the number of distinct authors rated after \(r\) ratings, \(N_{\text{asy}}\) represents the asymptotic number of authors a rater typically engages with, and \(\tau\) is a characteristic scale describing how rapidly a rater spreads their activity across authors.
To fit the saturation curve to the data, we used non-linear least squares optimization implemented via the \texttt{curve\_fit} function from the SciPy library~\cite{virtanen2020scipy}, which estimates the parameters \(N_{\text{asy}}\) and \(\tau\) by minimizing the sum of squared differences between the observed and predicted number of authors. We find that the characteristic scale \(\tau\) is substantially smaller in the observed data (\(\tau = 632\)) than in the randomized baseline (\(\tau = 1080\)), indicating that raters reach their asymptotic number of authors more quickly than expected by chance. 

Similarly, the asymptotic number of authors is lower in the observed data (\(N_{\text{asy}} = 427\)) compared to the randomized baseline (\(N_{\text{asy}} = 970\)), reflecting the overall extent of engagement.
The $CI_{95\%}$ for these parameter estimates, computed from the standard errors derived from the covariance matrix of the fitted parameters and the $t$-distribution, are as follows:
\[
\begin{aligned}
\text{Observed:} & \quad N_{\text{asy}} = 427.9 \pm 1.0, \quad 
\tau = 632.7 \pm 2.0 \\[1mm]
\text{Random:} & \quad N_{\text{asy}} = 970.0 \pm 0.7, \quad 
\tau = 1080.1 \pm 1.0
\end{aligned}
\]
The non-overlapping confidence intervals show that the observed and randomized data are clearly distinct, reinforcing that the observed behavior is not random.

\subsection{Polarization}
\begin{figure}[h]
  \centering
  \includegraphics[width=1\linewidth]{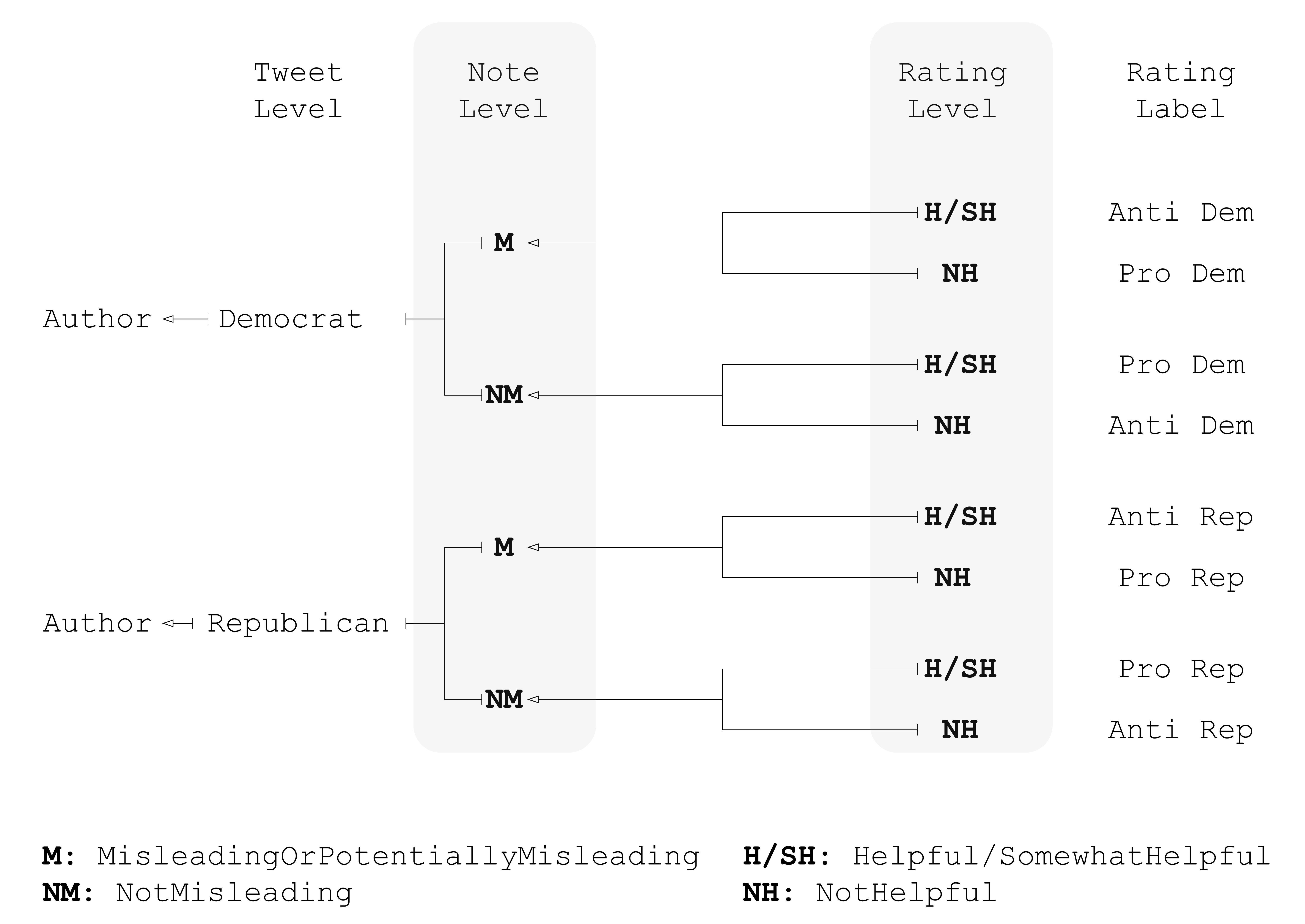}
  \caption{Schema of the evaluation pipeline. An author publishes a tweet and is classified as Democrat or Republican. A noter then produces a contextual note, labeled as \texttt{MisleadingOrPotentiallyMisleading} (M) or \texttt{NotMisleading} (NM). Finally, a rater evaluates the note by expressing agreement or disagreement, operationalized as \texttt{Helpful} or  \texttt{SomewhatHelpful} (H/SH) versus \texttt{NotHelpful} (NH). Each rating is subsequently interpreted as a political signal (pro/anti Democrat or Republican) by jointly considering the author’s affiliation, the type of note, and the rater’s judgment. }
  \label{schema}
\end{figure}

At this point, in order to study the association of hyper-activity with political bias, our goal is to estimate the political leaning of each rater within the CNs system.
On social media, political leaning inference is commonly framed as an indirect classification problem, in which users’ ideological positions are inferred from observable behavioral signals rather than self-reported affiliations. This is typically achieved by linking users to politically labeled entities, such as partisan media sources or political actors, and aggregating their interaction patterns with this content. Under the assumption that engagement reflects underlying political preferences, these interaction-based signals provide a proxy for users’ latent ideological orientation \cite{morales2015measuring,garimella2018quantifying,cinelli2021echo}.
To interpret each rating as a potential political signal, we combine information about the political leaning of the tweet author, the type of note, and the rating given. Our analysis focuses exclusively on the sample of tweets whose authors have a clearly identified political affiliation (Democrat or Republican), as defined in Section~\ref{subsec:methods}. 
As illustrated in Figure \ref{schema}, rating a note as \texttt{Helpful} when it flags a tweet as \texttt{MisleadingOrPotentiallyMisleading} can be interpreted as a signal of disagreement with the political stance of the tweet’s author. Therefore, we treat it as an \texttt{Anti} signal. Correspondingly, rating a note as \texttt{NotHelpful} can be interpreted as a \texttt{Pro} signal when the note flags a tweet as \texttt{MisleadingOrPotentiallyMisleading}. If the note instead classifies the tweet as \texttt{NotMisleading}, then a \texttt{Helpful} rating is treated as a \texttt{Pro} signal and a \texttt{NotHelpful} rating as an \texttt{Anti} signal.

Instead of analyzing a single signal, we aggregate multiple ratings to estimate systematic political biases in each rater’s behavior. For robustness, we restrict the analysis to users with at least $30$ ratings, ensuring a sufficiently large sample to reliably assign a political leaning label.
Formally, we define the political leaning of a rater as a real-valued measure $L$ taking values in $[-1, 1]$, given by
\begin{equation}
L = \frac{(a_{\text{Dem}} - p_{\text{Dem}}) + (p_{\text{Rep}} - a_{\text{Rep}})}
{\sum_{x \in \{\text{Dem}, \text{Rep}\}}(a_x + p_x)},
\end{equation}
where $a_x$ and $p_x$ denote, respectively, the number of \emph{anti} and \emph{pro} ratings expressed toward CNs supporting content produced by accounts labeled as either Democratic or Republican.
Positive values of $L$ indicate a Republican-leaning behavioral pattern, while negative values indicate a Democratic-leaning pattern. Values close to zero correspond to more balanced or weakly biased behavior.  Importantly, $L$ reflects observed interaction patterns within CNs rather than explicit ideological identification. In other words, it does not measure a rater's declared political beliefs, but acts as a \emph{proxy for systematic political bias} in their behavior, capturing tendencies to consistently defend authors of a particular political party while always attacking the opposing side.

The aggregated distribution of political leaning $L$, across all contributors, follows an approximately unimodal shape with a slight shift toward negative values, indicating a Democratic-leaning tendency. The descriptive statistics of $L$ are as follows: the mean is $\mu_L = -0.08$, the median is $M_L = -0.11$, the standard deviation is $\sigma_L = 0.69$, and the skewness is $\gamma_L = 0.19$. The skewness $\gamma_L$ was calculated using the \texttt{.skew()} method in Pandas, which returns the Fisher-Pearson coefficient of skewness, measuring the degree of asymmetry of the distribution: positive values indicate a longer right tail, while negative values indicate a longer left tail. In this case, the slightly positive $\gamma_L$ confirms that the distribution is nearly symmetric. At the aggregate level, this suggests a relatively balanced participation of users with different political orientations.

\begin{figure}[h]
  \centering
  \includegraphics[width=0.8\linewidth]{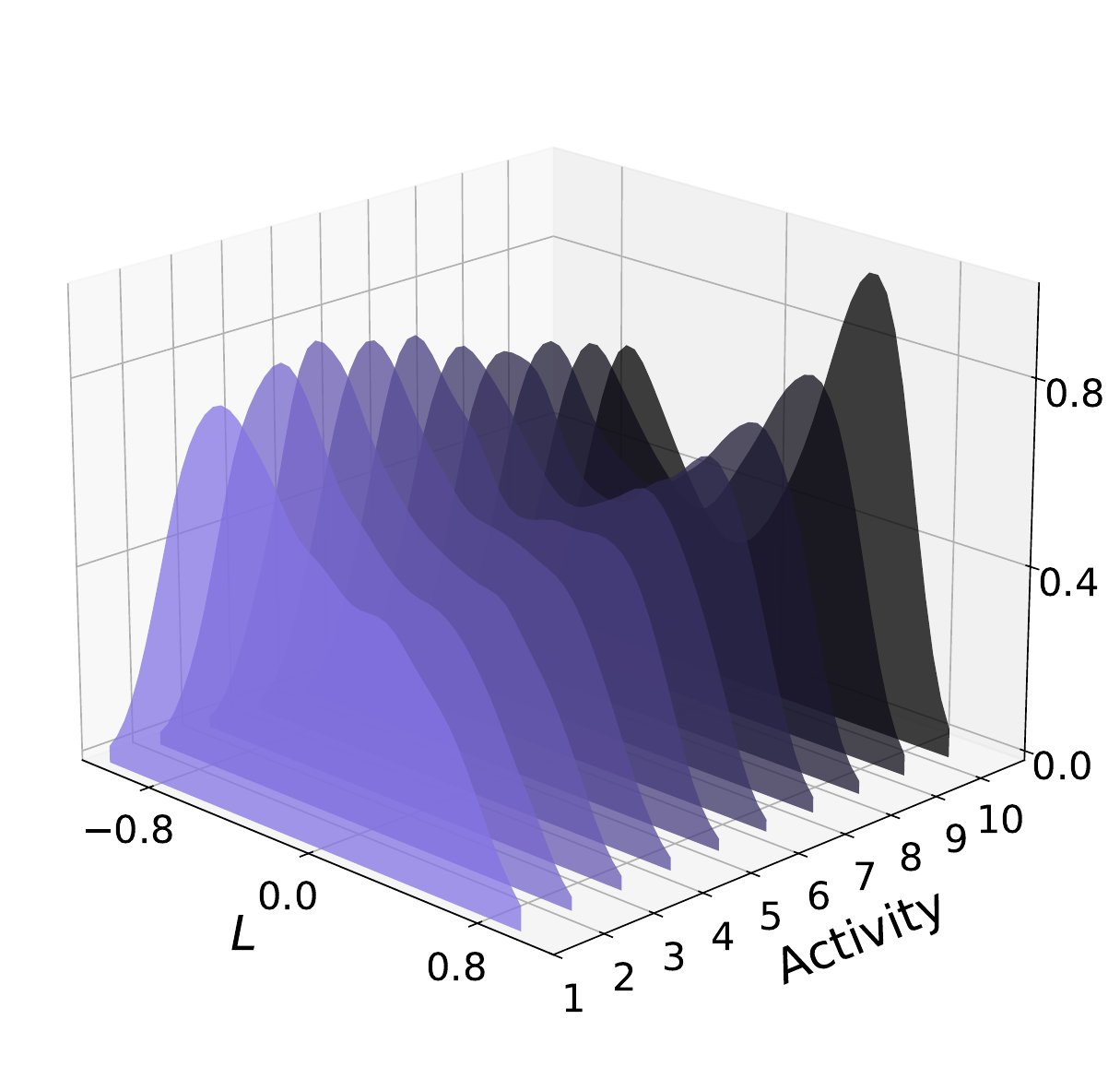}
  \caption{Distribution of political leaning $L$ across rating activity deciles. Negative (positive) values of $L$ indicate a preference toward rating notes associated with tweets by Democrat (Republican) accounts favorably. Polarization increases with user activity, indicating that the most active users tend to be the most politically polarized, and therefore engage in rating in a more biased manner.}
  \label{activ_pol}
\end{figure}

Motivated by the strong inequality and selectivity in rating activity documented in the previous section, we classify each user based on the total number of ratings they have provided on the CNs platform within the politically labeled sample. This allows us to investigate whether the more active users, responsible for a disproportionate share of ratings, also exhibit observable differences in political behavior. Specifically, we group users by activity deciles, with each decile representing 10\% of the rater population ordered by increasing activity. The resulting distributions of $L$ shown in Figure~\ref{activ_pol} reveal that contributors in the top decile of activity exhibit substantially more polarized behavior, with a clear shift toward positive values of $L$, indicative of a predominantly Republican-leaning pattern.
For each decile, we fit a Gaussian Mixture Model (GMM) with two components to the data, in order to identify potential subpopulations. 
A GMM assumes that the observed distribution can be represented as a weighted combination of multiple Gaussian distributions. 
By fitting the model, we estimate the mean ($\mu$), standard deviation ($\sigma$), and weight of each component. 
Using these parameters, we then compute \textit{Ashman's $D$}, a measure of bimodality that quantifies how well-separated two peaks in the distribution are:
\[
D = \frac{|\mu_1 - \mu_2|}{\sqrt{\frac{\sigma_1^2 + \sigma_2^2}{2}}}
\]
where $\mu_1$ and $\mu_2$ are the means of the two Gaussian components, and $\sigma_1$ and $\sigma_2$ are their standard deviations. 
Values of $D > 2$ indicate a bimodal distribution, with higher values corresponding to a stronger separation between the modes.

The Ashman’s $D$ values for selected activity deciles are 2.80, 2.73, 2.82, and 2.86 for deciles 1 to 4, and 3.63 and 4.17 for deciles 9 and 10. These results confirm an upward trend, indicating that the most active users exhibit more polarized behavior, whereas less active users display more mixed political ratings. This pattern suggests that increased activity is associated with a stronger concentration of users in distinct political positions, thereby reinforcing the bimodality of the distribution.

\subsection{Counterfactual analysis of notes emergence}
Given the presence of highly active polarized contributors identified in previous sections, a natural question is whether the statuses assigned by the Community Notes algorithm are particularly sensitive to their participation in the scoring process.
To address this question, we employ a counterfactual experimental framework that operates directly on the Community Notes production pipeline. In this context, a counterfactual refers to a hypothetical scenario in which the observable output of the system is recomputed after selectively perturbing a specific component of the input, while keeping all other elements unchanged. This design allows us to recompute~\cite{opensourcecode} the note statuses that the platform would have produced under alternative rating scenarios, rather than relying on proxy measures or theoretical assumptions. 

Specifically, we keep the full set of submitted notes and the consensus algorithm fixed, and construct alternative rating scenarios by modifying the composition of raters.
We modify only two data sources used as input by the algorithm: \textit{ratings} and \textit{noteStatusHistory} as explained in Section \ref{subsec:methods}. 
In the \textit{ratings} file, we progressively remove all ratings submitted by the most active users, excluding the top 10, 100, 200, 500, 1,000, 5,000, 10,000 raters.
These raters represent an extremely small fraction of contributors population, ranging from $7.69 \times 10^{-6} $ \% to $7.69 \times 10^{-3} $ \%. 
As for the \textit{noteStatusHistory}, we remove the fields that prevent the notes from being rescored (\ie being eligible for a change in status as computed by the algorithm), as in production X uses a locking mechanism that prevents rescoring for notes older than 14 days to reduce computational overhead.
Modifying these fields does not alter the scoring algorithm itself, but ensures that all notes are eligible for rescoring under each counterfactual experiment.
Considering simulations that remove increasing numbers of the most active users, our objective is to measure the stability of the note emergence by assessing the consistency of note statuses across counterfactual scenarios.
We define the baseline scenario as one in which all observed ratings remain unchanged ($0$ raters are removed).

\begin{figure}[h]
  \centering
  \includegraphics[width=1\linewidth]{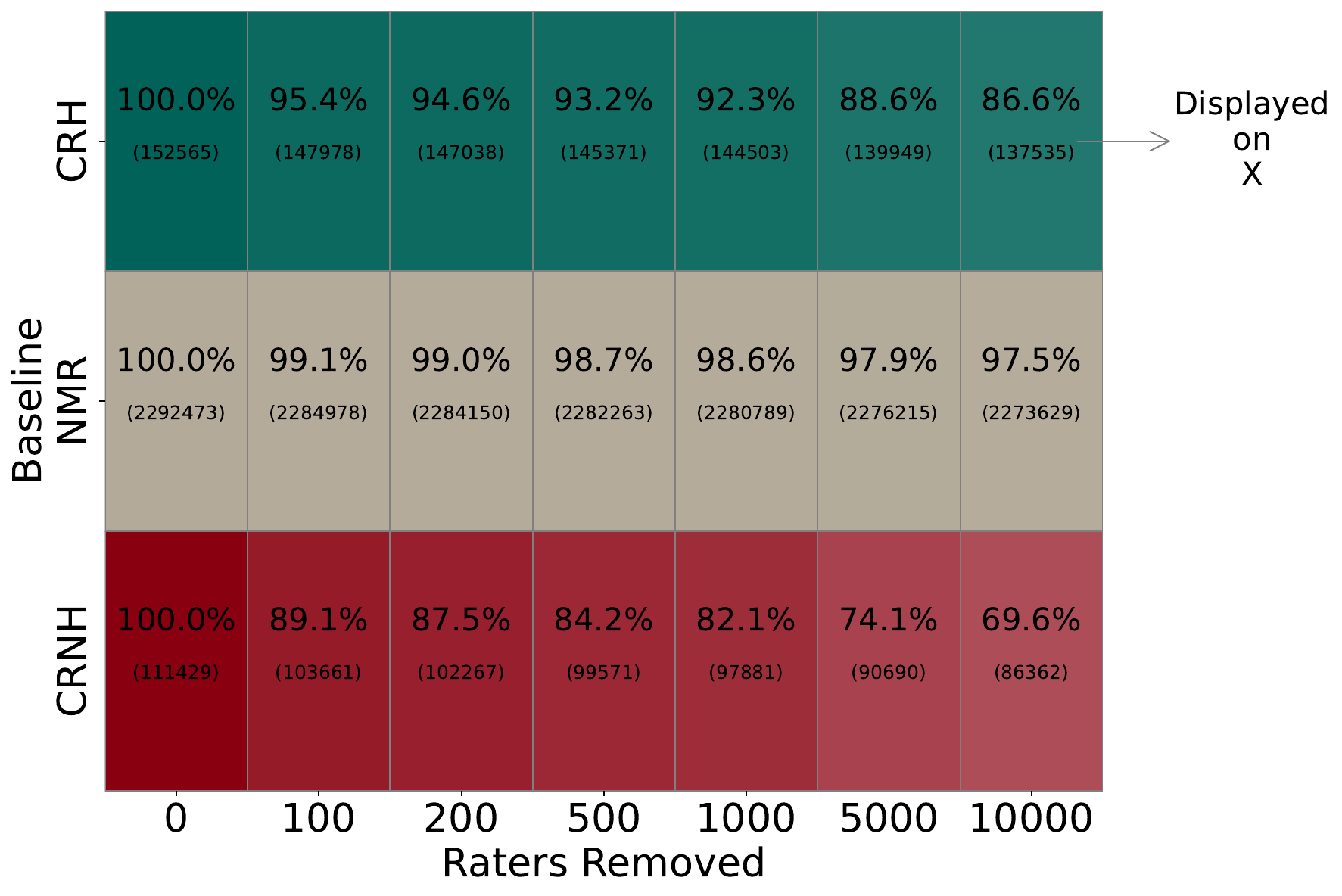}
\caption{
Heatmap illustrating note status under counterfactual simulations. 
Each row corresponds to a possible note status (NMR = \texttt{NeedsMoreRatings}, CRH = \texttt{CurrentlyRatedHelpful}, CRNH = \texttt{CurrentlyRatedNotHelpful}), while each column represents a counterfactual simulation obtained by removing a certain number of top raters.
Each cell reports the number of notes (in parentheses), and the Jaccard similarity (as a percentage) between the two sets of notes with that status in the baseline and simulation cases. 
As shown by the arrow, only notes in CRH are displayed to all X users.
}
\label{fig:emergence}
\end{figure}

To this end, we quantify the similarity between the sets of note statuses produced in each simulation to those observed in the baseline scenario using the Jaccard Similarity, defined as follows:
\[
J(B_{s}, S_{i,s}) = \frac{|N_{B_{s}} \cap N_{S_{i,s}}|}{|N_{S_{s}} \cup N_{S_{i,s}}|}
\]
where $N_{S_{I}}$ and $N_{S_{II}}$ denote the sets of notes, with a specific status $s$, in the baseline and the simulation with the removal of $i$ raters, respectively. Here, $|N_{B_{s}} \cap N_{S_{i,s}}|$ represents the number of notes with a status $s$ in both baseline and simulation, while $|N_{B_{s}} \cup N_{S_{i,s}}|$ is the number of notes with that status in at least one of the two.
As shown in Figure~\ref{fig:emergence}, for each simulation we calculated the Jaccard similarity between the note statuses and the baseline case, where no users were removed.

The heatmap shows that even when removing only the 10 most active users, 4,587 of the 152,565 notes marked as CRH (i.e., those visible to all users on X as illustrated in the example in Figure~\ref{fig:crhnotes_fun}) are no longer visible, because the algorithm no longer detects sufficient consensus, while 7,768 notes marked as CRNH change and are no longer in that status. Removing 10,000 users (approximately 0.007\% of the raters) causes the Jaccard similarity of the statuses to drop up to 69\% in the CRNH case, highlighting the sensitivity of the decisions computed by the algorithm in the baseline case. Furthermore, while changes in CRH statuses may appear more consequential, as they determine which notes ultimately are displayed, changes in CRNH are equally important. Since CRNH notes are collapsed by default by the system and require explicit user action to be viewed, this status substantially reduces a note’s visibility and may indirectly affect downstream consensus formation.

The persistence of this instability across all statuses and simulations underscores the disproportionate influence of a small group of hyperactive raters on algorithmic outcomes.

\section{Conclusions}
In this paper, we provided a large-scale empirical characterization of CNs as a community-based contextualization and fact-checking system. Leveraging the full public history of notes and ratings from 2021 to 2025 and accessing to the consensus algorithm used in production, we examined how participation patterns and political bias shape which notes emerge and become visible on X.

Our analysis shows that CNs replicate key structural dynamics typically observed in social media. Rating activity is highly concentrated, with a small minority of raters responsible for a disproportionate share of ratings: in particular, 80\% of all ratings in the system are produced by just 20\% of raters. Moreover, as the number of X users participating in Community Notes has grown over time, the rating activity has become increasingly concentrated among a shrinking core of highly active contributors, each focused on a limited set of authors. 
Furthermore, we observe that the most active raters are also the most susceptible to political bias. As their activity increases, the distribution of political leanings becomes progressively more polarized, indicating that highly active raters are disproportionately involved in partisan dynamics. Consequently, the users who exert the greatest influence on the rating process do not mirror the broader user base, but instead hold more extreme and asymmetric political views. This has the potential to affect the neutrality and balance of the Community Notes evaluation process.
Our study also shows that these participation asymmetries interact with the consensus algorithm that determines note visibility. Although the algorithm is explicitly designed to require agreement across politically heterogeneous users, the disproportionate influence of a small subset of highly active and polarized raters can affect algorithmic outputs by both inhibiting and stimulating the emergence of notes. Indeed, by applying the consensus algorithm that governs the emergence of notes, we investigated what would happen if we excluded small fractions of highly active users—ranging from $7.69 \times 10^{-6}\%$ to $7.69 \times 10^{-3}\%$ of the total population of raters.
Through these counterfactual simulations, we showed that even removing a tiny subset of highly active raters can significantly influence the set of notes that emerge.

Our analysis has several limitations. While we leverage the official open-source implementation of the Community Notes consensus algorithm, we cannot fully reproduce all production-level constraints, such as real-time batching, internal locking mechanisms, or undocumented platform-specific heuristics. These factors may affect the absolute timing or visibility of individual notes, but are unlikely to alter the qualitative sensitivity of the system to participation heterogeneity, which is the focus of our counterfactual analysis. 
In addition, political leaning is inferred from observable rating behavior rather than from self-reported ideology. This approach captures behavioral polarization within the rating process, but should be interpreted as a proxy rather than a direct measure of users’ political bias. 
Finally, our counterfactual scenarios perturb raters' composition while holding note content and algorithmic logic fixed. As a result, our findings isolate the structural role of participation inequality, but do not account for potential adaptive responses by users or by the platform over time.

Future studies could examine whether a small fraction of highly active users creates a structural bottleneck in the note publication process. Fact-checking interventions are inherently time-sensitive, and the concentration of activity among a limited set of raters may slow down consensus formation, delaying note visibility. Given that most tweets are consumed within approximately $\sim72$ hours, even modest delays can substantially reduce the effectiveness of corrective interventions. Another aspect worth exploring is how the consensus algorithm could be modified to improve robustness to participation asymmetries, for example, by capping individual activity levels or dynamically down-weighting hyperactive users.

While Community Notes are often praised for their transparency, scalability, and perceived legitimacy, our results show that the system is deeply shaped by participation inequalities typical of online platforms. Rather than producing a genuinely collective and democratic process, epistemic authority results in being concentrated in the hands of a small, highly active subset of users. In line with recent work questioning the sufficiency of community-based corrections alone \cite{vraga}, our findings suggest that the “wisdom of the crowd” is difficult to realize in practice. Community-based fact-checking exhibits structural limitations that must be explicitly addressed if it is to function as a reliable complement or alternative to professional fact-checking systems.

\bibliographystyle{plain}
\bibliography{aaai2026}

@article{del2016spreading,
  title={The spreading of misinformation online},
  author={Del Vicario, Michela and Bessi, Alessandro and Zollo, Fabiana and Petroni, Fabio and Scala, Antonio and Caldarelli, Guido and Stanley, H Eugene and Quattrociocchi, Walter},
  journal={Proceedings of the national academy of Sciences},
  volume={113},
  number={3},
  pages={554--559},
  year={2016},
  publisher={National Academy of Sciences}
}

@article{zollo2017debunking,
  title={Debunking in a world of tribes},
  author={Zollo, Fabiana and Bessi, Alessandro and Del Vicario, Michela and Scala, Antonio and Caldarelli, Guido and Shekhtman, Louis and Havlin, Shlomo and Quattrociocchi, Walter},
  journal={PloS one},
  volume={12},
  number={7},
  pages={e0181821},
  year={2017},
  publisher={Public Library of Science San Francisco, CA USA}
}

@article{cinelli2020covid,
  title={The COVID-19 social media infodemic},
  author={Cinelli, Matteo and Quattrociocchi, Walter and Galeazzi, Alessandro and Valensise, Carlo Michele and Brugnoli, Emanuele and Schmidt, Ana Lucia and Zola, Paola and Zollo, Fabiana and Scala, Antonio},
  journal={Scientific reports},
  volume={10},
  number={1},
  pages={16598},
  year={2020},
  publisher={Nature Publishing Group UK London}
}

@article{del2017modeling,
  title={Modeling confirmation bias and polarization},
  author={Del Vicario, Michela and Scala, Antonio and Caldarelli, Guido and Stanley, H Eugene and Quattrociocchi, Walter},
  journal={Scientific reports},
  volume={7},
  number={1},
  pages={40391},
  year={2017},
  publisher={Nature Publishing Group UK London}
}

@book{pareto1964cours,
  title={Cours d'{\'e}conomie politique},
  author={Pareto, Vilfredo},
  volume={1},
  year={1964},
  publisher={Librairie Droz}
}

@inproceedings{saeed2022crowdsourced,
  title={Crowdsourced fact-checking at Twitter: How does the crowd compare with experts?},
  author={Saeed, Mohammed and Traub, Nicolas and Nicolas, Maelle and Demartini, Gianluca and Papotti, Paolo},
  booktitle={Proceedings of the 31st ACM international conference on information \& knowledge management},
  pages={1736--1746},
  year={2022}
}

@article{gini1921measurement,
  title={Measurement of inequality of incomes},
  author={Gini, Corrado},
  journal={The economic journal},
  volume={31},
  number={121},
  pages={124--125},
  year={1921},
  publisher={Oxford University Press Oxford, UK}
}

@inproceedings{nogara2022disinformation,
  title={The disinformation dozen: An exploratory analysis of covid-19 disinformation proliferation on twitter},
  author={Nogara, Gianluca and Vishnuprasad, Padinjaredath Suresh and Cardoso, Felipe and Ayoub, Omran and Giordano, Silvia and Luceri, Luca},
  booktitle={Proceedings of the 14th ACM web science conference 2022},
  pages={348--358},
  year={2022}
}

@inproceedings{dixon2018measuring,
  title={Measuring and mitigating unintended bias in text classification},
  author={Dixon, Lucas and Li, John and Sorensen, Jeffrey and Thain, Nithum and Vasserman, Lucy},
  booktitle={Proceedings of the 2018 AAAI/ACM Conference on AI, Ethics, and Society},
  pages={67--73},
  year={2018}
}

@inproceedings{pilarski2024community,
  title={Community notes vs. snoping: how the crowd selects fact-checking targets on social media},
  author={Pilarski, Moritz and Solovev, Kirill Olegovich and Pr{\"o}llochs, Nicolas},
  booktitle={Proceedings of the International AAAI Conference on Web and Social Media},
  volume={18},
  pages={1262--1275},
  year={2024}
}

@article{drolsbach2024community,
  title={Community notes increase trust in fact-checking on social media},
  author={Drolsbach, Chiara Patricia and Solovev, Kirill and Pr{\"o}llochs, Nicolas},
  journal={PNAS nexus},
  volume={3},
  number={7},
  pages={pgae217},
  year={2024},
  publisher={Oxford University Press US}
}

@article{allen2021scaling,
  title={Scaling up fact-checking using the wisdom of crowds},
  author={Allen, Jennifer and Arechar, Antonio A and Pennycook, Gordon and Rand, David G},
  journal={Science advances},
  volume={7},
  number={36},
  pages={eabf4393},
  year={2021},
  publisher={American Association for the Advancement of Science}
}

@article{li2025scaling,
  title={Scaling Human Judgment in Community Notes with LLMs},
  author={Li, Haiwen and De, Soham and Revel, Manon and Haupt, Andreas and Miller, Brad and Coleman, Keith and Baxter, Jay and Saveski, Martin and Bakker, Michiel A},
  journal={arXiv preprint arXiv:2506.24118},
  year={2025}
}

@article{renault2025republicans,
  title={Republicans are flagged more often than Democrats for sharing misinformation on X’s Community Notes},
  author={Renault, Thomas and Mosleh, Mohsen and Rand, David G},
  journal={Proceedings of the National Academy of Sciences},
  volume={122},
  number={25},
  pages={e2502053122},
  year={2025},
  publisher={National Academy of Sciences}
}

@article{chuai2023roll,
  title={The roll-out of community notes did not reduce engagement with misinformation on Twitter},
  author={Chuai, Yuwei and Tian, Haoye and Pr{\"o}llochs, Nicolas and Lenzini, Gabriele},
  journal={arXiv e-prints},
  pages={arXiv--2307},
  year={2023}
}

@article{slaughter2025community,
  title={Community notes moderate engagement with and diffusion of false information online},
  author={Slaughter, Isaac and Peytavin, Axel and Ugander, Johan and Saveski, Martin},
  journal={arXiv preprint arXiv:2502.13322},
  year={2025}
}

@article{kangur2024checks,
  title={Who Checks the Checkers? Exploring Source Credibility in Twitter's Community Notes},
  author={Kangur, Uku and Chakraborty, Roshni and Sharma, Rajesh},
  journal={arXiv preprint arXiv:2406.12444},
  year={2024}
}

@article{borenstein2025can,
  title={Can Community Notes Replace Professional Fact-Checkers?},
  author={Borenstein, Nadav and Warren, Greta and Elliott, Desmond and Augenstein, Isabelle},
  journal={arXiv preprint arXiv:2502.14132},
  year={2025}
}

@article{matamoros2025importance,
  title={The importance of centering harm in data infrastructures for ‘soft moderation’: X’s Community Notes as a case study},
  author={Matamoros-Fern{\'a}ndez, Ariadna and Jude, Nadia},
  journal={New Media \& Society},
  volume={27},
  number={4},
  pages={1986--2011},
  year={2025},
  publisher={SAGE Publications Sage UK: London, England}
}

@article{sharot2020people,
  title={How people decide what they want to know},
  author={Sharot, Tali and Sunstein, Cass R},
  journal={Nature Human Behaviour},
  volume={4},
  number={1},
  pages={14--19},
  year={2020},
  publisher={Nature Publishing Group UK London}
}

@inproceedings{bakshy2012role,
  title={The role of social networks in information diffusion},
  author={Bakshy, Eytan and Rosenn, Itamar and Marlow, Cameron and Adamic, Lada},
  booktitle={Proceedings of the 21st international conference on World Wide Web},
  pages={519--528},
  year={2012}
}

@article{lazer2018science,
  title={The science of fake news},
  author={Lazer, David MJ and Baum, Matthew A and Benkler, Yochai and Berinsky, Adam J and Greenhill, Kelly M and Menczer, Filippo and Metzger, Miriam J and Nyhan, Brendan and Pennycook, Gordon and Rothschild, David and others},
  journal={Science},
  volume={359},
  number={6380},
  pages={1094--1096},
  year={2018},
  publisher={American Association for the Advancement of Science}
}

@article{
theSpread,
author = {Soroush Vosoughi  and Deb Roy  and Sinan Aral },
title = {The spread of true and false news online},
journal = {Science},
volume = {359},
number = {6380},
pages = {1146-1151},
year = {2018},
doi = {10.1126/science.aap9559},
URL = {https://www.science.org/doi/abs/10.1126/science.aap9559},
eprint = {https://www.science.org/doi/pdf/10.1126/science.aap9559}}

@article{cinelli2021echo,
  title={The echo chamber effect on social media},
  author={Cinelli, Matteo and De Francisci Morales, Gianmarco and Galeazzi, Alessandro and Quattrociocchi, Walter and Starnini, Michele},
  journal={Proceedings of the national academy of sciences},
  volume={118},
  number={9},
  pages={e2023301118},
  year={2021},
  publisher={National Academy of Sciences}
}

@misc{fbMoreSpeech,
  author       = {{Meta}},
  title = {{M}ore {S}peech and {F}ewer {M}istakes --- about.fb.com},
  howpublished = {\url{https://about.fb.com/news/2025/01/meta-more-speech-fewer-mistakes/}},
  year         = {2025},
  note         = {Accessed: 22 Oct 2025}
}

@misc{opensourcecode,
  author       = {{Twitter}},
  title = {Open-source code --- communitynotes.x.com},
  howpublished = {\url{https://communitynotes.x.com/guide/en/under-the-hood/note-ranking-code}},
year=2022,
  note         = {Accessed: 09 Jan 2026}
}

@inproceedings{allen2022birds,
  title={Birds of a feather don’t fact-check each other: Partisanship and the evaluation of news in Twitter’s Birdwatch crowdsourced fact-checking program},
  author={Allen, Jennifer and Martel, Cameron and Rand, David G},
  booktitle={Proceedings of the 2022 CHI conference on human factors in computing systems},
  pages={1--19},
  year={2022}
}

@misc{xIntroduction,
  author       = {{X Community Notes}},
  year = {n.d.},
  title        = {Community Notes: A Collaborative Way to Add Helpful Context to Posts and Keep People Better Informed},
  howpublished = {\url{https://communitynotes.x.com/guide/en/about/introduction}},
  note         = {Accessed: 22 Oct 2025}
}

@ARTICLE{nudo_size,
  author={Nudo, Jacopo and Cinelli, Matteo and Baronchelli, Andrea and Quattrociocchi, Walter},
  journal={IEEE Transactions on Computational Social Systems}, 
  title={From Niche to Mainstream: Community Size and Engagement in Social Media Conversations}, 
  year={2025},
  volume={},
  number={},
  pages={1-8},
  doi={10.1109/TCSS.2025.3604221}}

@article{bakshy2015exposure,
  title={Exposure to ideologically diverse news and opinion on Facebook},
  author={Bakshy, Eytan and Messing, Solomon and Adamic, Lada A},
  journal={Science},
  volume={348},
  number={6239},
  pages={1130--1132},
  year={2015},
  publisher={American Association for the Advancement of Science}
}

@article{cinelli2025infodemic,
  title={Infodemic Versus Viral Information Spread: Key Differences and Open Challenges},
  author={Cinelli, Matteo and Gesualdo, Francesco and others},
  journal={JMIR infodemiology},
  volume={5},
  number={1},
  pages={e57455},
  year={2025},
  publisher={JMIR Publications Inc., Toronto, Canada}
}

@inproceedings{de2025supernotes,
  title={Supernotes: Driving consensus in crowd-sourced fact-checking},
  author={De, Soham and Bakker, Michiel A and Baxter, Jay and Saveski, Martin},
  booktitle={Proceedings of the ACM on Web Conference 2025},
  pages={3751--3761},
  year={2025}
}

@article{wojcik2022birdwatch,
  title={Birdwatch: Crowd wisdom and bridging algorithms can inform understanding and reduce the spread of misinformation},
  author={Wojcik, Stefan and Hilgard, Sophie and Judd, Nick and Mocanu, Delia and Ragain, Stephen and Hunzaker, MB and Coleman, Keith and Baxter, Jay},
  journal={arXiv preprint arXiv:2210.15723},
  year={2022}
}

@article{avalle2024persistent,
  title={Persistent interaction patterns across social media platforms and over time},
  author={Avalle, Michele and Di Marco, Niccol{\`o} and Etta, Gabriele and Sangiorgio, Emanuele and Alipour, Shayan and Bonetti, Anita and Alvisi, Lorenzo and Scala, Antonio and Baronchelli, Andrea and Cinelli, Matteo and others},
  journal={Nature},
  volume={628},
  number={8008},
  pages={582--589},
  year={2024},
  publisher={Nature Publishing Group UK London}
}

@article{del2016echo,
  title={Echo chambers: Emotional contagion and group polarization on facebook},
  author={Del Vicario, Michela and Vivaldo, Gianna and Bessi, Alessandro and Zollo, Fabiana and Scala, Antonio and Caldarelli, Guido and Quattrociocchi, Walter},
  journal={Scientific reports},
  volume={6},
  number={1},
  pages={37825},
  year={2016},
  publisher={Nature Publishing Group UK London}
}

@article{
vraga,
    author = {Emily K. Vraga },
    title = {Understanding the strengths and limitations of community-based responses to misinformation},
    journal = {Proceedings of the National Academy of Sciences},
    volume = {122},
    number = {48},
    pages = {e2524004122},
    year = {2025},
    doi = {10.1073/pnas.2524004122},
    URL = {https://www.pnas.org/doi/abs/10.1073/pnas.2524004122},
    eprint = {https://www.pnas.org/doi/pdf/10.1073/pnas.2524004122}}

@article{gonzalez2023asymmetric,
  title={Asymmetric ideological segregation in exposure to political news on Facebook},
  author={Gonz{\'a}lez-Bail{\'o}n, Sandra and Lazer, David and Barber{\'a}, Pablo and Zhang, Meiqing and Allcott, Hunt and Brown, Taylor and Crespo-Tenorio, Adriana and Freelon, Deen and Gentzkow, Matthew and Guess, Andrew M and others},
  journal={Science},
  volume={381},
  number={6656},
  pages={392--398},
  year={2023},
  publisher={American Association for the Advancement of Science}
}

@article{truong2025community,
  title={Community Notes are Vulnerable to Rater Bias and Manipulation},
  author={Truong, Bao Tran and Wu, Siqi and Flammini, Alessandro and Menczer, Filippo and Stewart, Alexander J},
  journal={arXiv preprint arXiv:2511.02615},
  year={2025}
}

@article{cota2019quantifying,
  title={Quantifying echo chamber effects in information spreading over political communication networks},
  author={Cota, Wesley and Ferreira, Silvio C and Pastor-Satorras, Romualdo and Starnini, Michele},
  journal={EPJ Data Science},
  volume={8},
  number={1},
  pages={1--13},
  year={2019},
  publisher={Springer}
}

@misc{binder2023communitynotes,
  title        = {Most users on X never see Community Notes correcting misinformation},
  author       = {Binder, Matt},
  howpublished = {\url{https://mashable.com/article/twitter-x-community-notes-misinformation-views-investigation}},
  year         = {2023},
  note         = {Mashable, accessed December 15, 2025}
}

@article{arjmandi2025threats,
  title={Threats to the sustainability of Community Notes on X},
  author={Arjmandi-Lari, Zahra and Mantzarlis, Alexios and Stafford, Tom},
  journal={arXiv preprint arXiv:2510.00650},
  year={2025}
}

@article{brashier2021timing,
  title={Timing matters when correcting fake news},
  author={Brashier, Nadia M and Pennycook, Gordon and Berinsky, Adam J and Rand, David G},
  journal={Proceedings of the National Academy of Sciences},
  volume={118},
  number={5},
  pages={e2020043118},
  year={2021},
  publisher={National Academy of Sciences}
}

@article{truong2025delayed,
  title={Delayed takedown of illegal content on social media makes moderation ineffective},
  author={Truong, Bao Tran and Kim, Sangyeon and Nogara, Gianluca and Verdolotti, Enrico and Sahneh, Erfan Samieyan and Saurwein, Florian and Just, Natascha and Luceri, Luca and Giordano, Silvia and Menczer, Filippo},
  journal={arXiv preprint arXiv:2502.08841},
  year={2025}
}

@article{mohammadi2025birdwatch,
  title={From Birdwatch to Community Notes, from Twitter to X: four years of community-based content moderation},
  author={Mohammadi, Saeedeh and Chinichian, Narges and Doyal, Hannah and Skutilova, Kristina and Cui, Hao and d'Errico, Michele and Grayson, Siobhan and Yasseri, Taha},
  journal={arXiv preprint arXiv:2510.09585},
  year={2025}
}

@article{kim2025differential,
  title={Differential impact from individual versus collective misinformation tagging on the diversity of Twitter (X) information engagement and mobility},
  author={Kim, Junsol and Wang, Zhao and Shi, Haohan and Ling, Hsin-Keng and Evans, James},
  journal={Nature Communications},
  volume={16},
  number={1},
  pages={973},
  year={2025},
  publisher={Nature Publishing Group UK London}
}

@article{martel2024crowds,
  title={Crowds can effectively identify misinformation at scale},
  author={Martel, Cameron and Allen, Jennifer and Pennycook, Gordon and Rand, David G},
  journal={Perspectives on Psychological Science},
  volume={19},
  number={2},
  pages={477--488},
  year={2024},
  publisher={Sage Publications Sage CA: Los Angeles, CA}
}

@article{la2024crowdsourced,
  title={Crowdsourced Fact-checking: Does It Actually Work?},
  author={La Barbera, David and Maddalena, Eddy and Soprano, Michael and Roitero, Kevin and Demartini, Gianluca and Ceolin, Davide and Spina, Damiano and Mizzaro, Stefano},
  journal={Information Processing \& Management},
  volume={61},
  number={5},
  pages={103792},
  year={2024},
  publisher={Elsevier}
}

@misc{metaIntroducingCommunity,
  author       = {{Meta}},
  title = {{I}ntroducing {C}ommunity {N}otes - {A}dding {C}ontext to {P}osts | {M}eta --- meta.com},
  howpublished = {\url{https://www.meta.com/technologies/community-notes/}},
  year         = {2025},
  note         = {Accessed: 22 Oct 2025}
}

@misc{apNewsCrowd-sourced,
	author = {Barbara Ortutay},
	title = {{M}eta to start testing crowd-sourced fact-checking, based on {X} example, next week --- apnews.com},
	howpublished = {\url{https://apnews.com/article/meta-fact-checks-community-notes-bb814cfc5e8d29a1ecc058f836de9580}},
	year = {2025},
	note = {[Accessed 12-01-2026]},
}

@misc{tiktokTestingFeature,
  author       = {{TikTok Newsroom}},
  title        = {{T}esting a new feature to enhance content on {T}ik{T}ok - {N}ewsroom | {T}ik{T}ok --- newsroom.tiktok.com},
  howpublished = {\url{https://newsroom.tiktok.com/footnotes}},
  year         = {2025},
  note         = {Accessed: 22 Oct 2025}
}

@inproceedings{prollochs2022community,
  title={Community-based fact-checking on Twitter’s Birdwatch platform},
  author={Pr{\"o}llochs, Nicolas},
  booktitle={Proceedings of the International AAAI Conference on Web and Social Media},
  volume={16},
  pages={794--805},
  year={2022}
}

@article{augenstein2025community,
  title={Community Moderation and the New Epistemology of Fact Checking on Social Media},
  author={Augenstein, Isabelle and Bakker, Michiel and Chakraborty, Tanmoy and Corney, David and Ferrara, Emilio and Gurevych, Iryna and Hale, Scott and Hovy, Eduard and Ji, Heng and Larraz, Irene and others},
  journal={arXiv preprint arXiv:2505.20067},
  year={2025}
}

@article{yasseri2023can,
  title={Can crowdsourcing rescue the social marketplace of ideas?},
  author={Yasseri, Taha and Menczer, Filippo},
  journal={Communications of the ACM},
  volume={66},
  number={9},
  pages={42--45},
  year={2023},
  publisher={ACM New York, NY, USA}
}

@misc{counterhate,
	author = {Imran Ahmed},
	title = {counterhate.com},
	howpublished = {\url{https://counterhate.com/wp-content/uploads/2024/10/CCDH.CommunityNotes.FINAL-30.10.pdf}},
	year = {2024},
	note = {[Accessed 22-10-2025]},
}

@article{garimella2018quantifying,
  title={Quantifying controversy on social media},
  author={Garimella, Kiran and Morales, Gianmarco De Francisci and Gionis, Aristides and Mathioudakis, Michael},
  journal={ACM Transactions on Social Computing},
  volume={1},
  number={1},
  pages={1--27},
  year={2018},
  publisher={ACM New York, NY, USA}
}

@article{virtanen2020scipy,
  title={SciPy 1.0: fundamental algorithms for scientific computing in Python},
  author={Virtanen, Pauli and Gommers, Ralf and Oliphant, Travis E and Haberland, Matt and Reddy, Tyler and Cournapeau, David and Burovski, Evgeni and Peterson, Pearu and Weckesser, Warren and Bright, Jonathan and others},
  journal={Nature methods},
  volume={17},
  number={3},
  pages={261--272},
  year={2020},
  publisher={Nature Publishing Group US New York}
}

@article{clauset2009power,
  title={Power-law distributions in empirical data},
  author={Clauset, Aaron and Shalizi, Cosma Rohilla and Newman, Mark EJ},
  journal={SIAM review},
  volume={51},
  number={4},
  pages={661--703},
  year={2009},
  publisher={SIAM}
}

@article{morales2015measuring,
  title={Measuring political polarization: Twitter shows the two sides of Venezuela},
  author={Morales, Alfredo Jose and Borondo, Javier and Losada, Juan Carlos and Benito, Rosa M},
  journal={Chaos: An Interdisciplinary Journal of Nonlinear Science},
  volume={25},
  number={3},
  year={2015},
  publisher={AIP Publishing}
}

@article{persily2020social,
  title={Social media and democracy: The state of the field, prospects for reform},
  author={Persily, Nathaniel and Tucker, Joshua A and Tucker, Joshua Aaron},
  year={2020},
  publisher={Cambridge University Press}
}

@inproceedings{perez2018automatic,
  title={Automatic detection of fake news},
  author={P{\'e}rez-Rosas, Ver{\'o}nica and Kleinberg, Bennett and Lefevre, Alexandra and Mihalcea, Rada},
  booktitle={Proceedings of the 27th international conference on computational linguistics},
  pages={3391--3401},
  year={2018}
}

@article{shu2017fake,
  title={Fake news detection on social media: A data mining perspective},
  author={Shu, Kai and Sliva, Amy and Wang, Suhang and Tang, Jiliang and Liu, Huan},
  journal={ACM SIGKDD explorations newsletter},
  volume={19},
  number={1},
  pages={22--36},
  year={2017},
  publisher={ACM New York, NY, USA}
}

\end{document}